\documentclass{vgtc}                          

\usepackage{enumitem}

\graphicspath{{figures/}{pictures/}{images/}{./}} 

\usepackage{times}                     

\usepackage{tabu}                      
\usepackage{booktabs}                  
\usepackage{lipsum}                    
\usepackage{mwe}                       

\usepackage{mathptmx}                  

\onlineid{0}

\vgtccategory{Research}

\vgtcinsertpkg

\title{XplainAct: Visualization for Personalized Intervention Insights}

\author{Yanming Zhang \thanks{e-mail: yanming.zhang@stonybrook.edu}\\ %
        \scriptsize Stony Brook University %
\and Krishnakumar Hegde\thanks{e-mail:khegde@cs.stonybrook.edu}\\ %
        \scriptsize Stony Brook University %
\and Klaus Mueller\thanks{e-mail: mueller@cs.stonybrook.edu}\\ %
        \scriptsize Stony Brook University}

\teaser{
  \centering
  \includegraphics[width=0.8\linewidth]{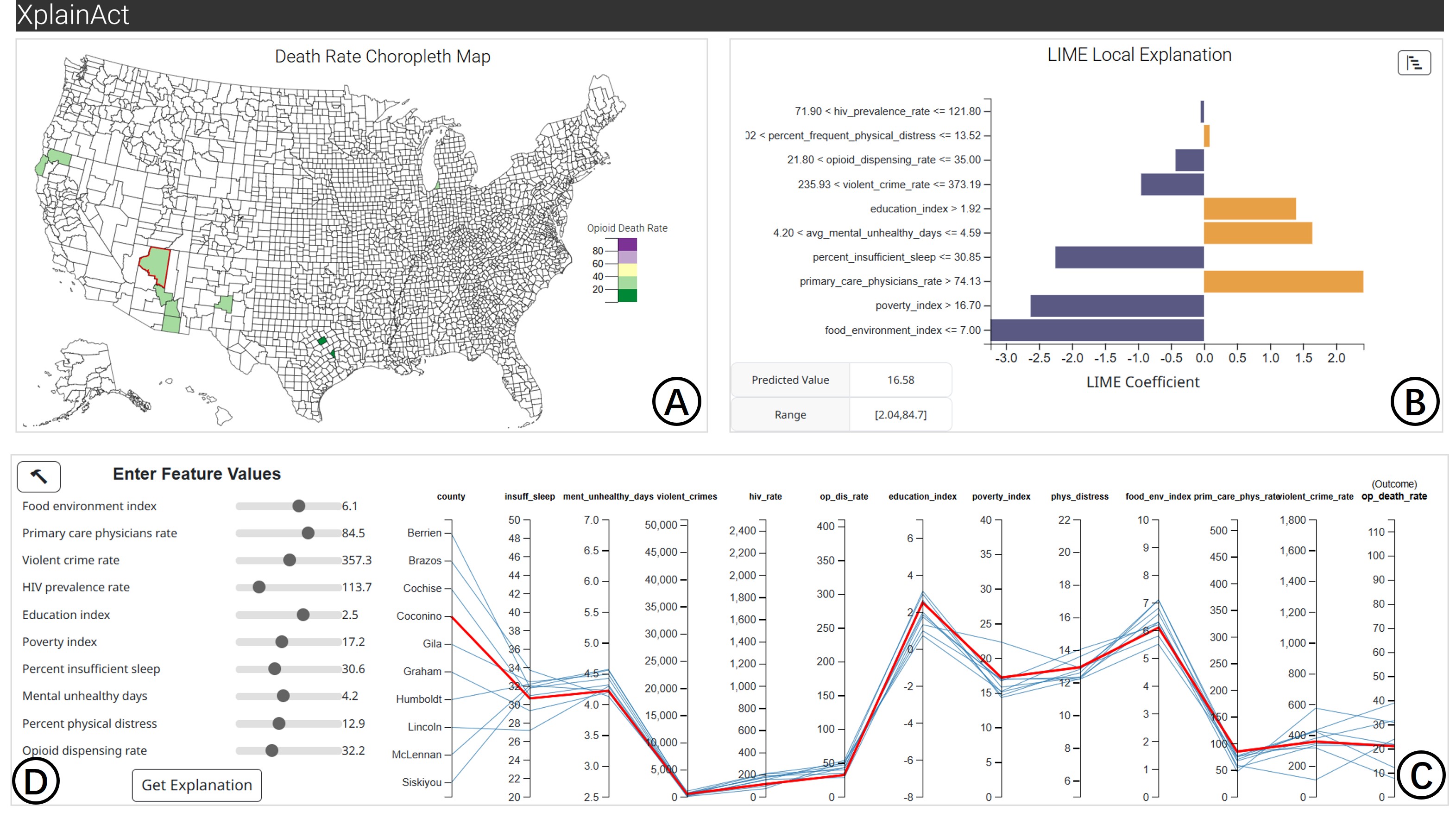}
  \vspace{-5pt}
  \caption{The XplainAct interface, illustrated here using the opioid dataset. (A) Choropleth map highlighting Coconino County, Arizona (bold red outline) and counties with similar opioid-related socioeconomic indicators with their respective opioid death rates colored according to the legend on the right; (B) Feature explanation panel using LIME to reveal the contribution of each socioeconomic factor to the opioid death rate in Coconino County; (C) Parallel coordinates plot comparing Coconino County’s multidimensional profile (red line) with its socioeconomic peers (blue lines) and highlighted on the choropleth map (A); and (D) Slider group for defining contextual similarity criteria to Coconino County’s profile.} 
  
  \label{fig:teaser}
}

\abstract{
        Causality helps people reason about and understand complex systems, particularly through what-if analyses that explore how interventions might alter outcomes. Although existing methods embrace causal reasoning using interventions and counterfactual analysis, they primarily focus on effects at the population level. These approaches often fall short in systems characterized by significant heterogeneity, where the impact of an intervention can vary widely across subgroups. To address this challenge, we present XplainAct, a visual analytics framework that supports
        simulating, explaining, and reasoning interventions at the individual level within subpopulations. We demonstrate the effectiveness of XplainAct through two case studies: investigating opioid-related deaths in epidemiology and analyzing voting inclinations in the presidential election.
} 


\begin{document}


\firstsection{Introduction}

\maketitle

The advances in machine learning and artificial intelligence in recent years have created a growing need for tools that can effectively support the understanding and modification of complex systems. Traditional analytical methods, which rely on correlation, merely observe how variables tend to change together. In contrast, causal reasoning goes a step further by determining whether changes in certain variables can causally influence others. Here, independent and dependent variables are often viewed as treatment and outcome variables, respectively, with the effect commonly quantified as the \textit{average treatment effect (ATE)}. 

The ATE is commonly used in observational studies. In standard causal inference settings, each individual receives only one treatment or exposure, meaning that only one outcome is observed, while the outcome under the alternative treatment—the counterfactual—remains unobserved. The ATE estimates the causal effect at the population level by conceptually treating individuals exposed to the alternative treatment as stand-ins for the unobserved counterfactuals. This estimand has proven effective in one-size-fits-all analyses, such as determining whether smoking causes cancer \cite{doll1950smoking}, sleep deprivation leads to cognitive impairment \cite{alhola2007sleep}, or anthropogenic greenhouse gas emissions drive global warming \cite{pachauri2014climate}.

Despite the usefulness of the ATE in uncovering general patterns, its averaging process can obscure critical variation in heterogeneous systems. A treatment that appears beneficial on average may be ineffective—or even harmful—for certain subgroups within the population. For instance, Manson et al. \cite{manson2013women} found that hormone therapy benefited younger postmenopausal women but posed risks for older women. Similarly, Kravitz et al. \cite{kravitz2004evidence} showed that the risk of gastrointestinal bleeding from aspirin use varied significantly depending on a patient’s age and history of peptic ulcers.

The examples above underscore the need to account for variations in individual responses to interventions. Yet, existing approaches face two key challenges. First, due to regulations and ethical concerns, it is often infeasible to apply certain interventions—such as assigning students to underperforming schools to study the effects of poor educational environments—thus limiting the use of randomized trials. Second, explainability is essential for building practitioner trust. Many AI-based personalization tools operate opaquely, and practitioners often distrust recommendations that lack a clear rationale. This poses a barrier to adoption in high-stakes domains like clinical medicine \cite{vallee2024envisioning}. These challenges underscore the need for systems that can personalize, explore, and interpret intervention effects—motivating the development of a visual analytics tool to support human-in-the-loop what-if analysis.
 
In this work, we propose XplainAct, a visual analytics framework designed to support the estimation and interpretation of individual-level intervention effects within subgroups. We demonstrate how XplainAct can recommend the most effective interventions to shift outcome values toward desired targets through two illustrative usage scenarios.

\section{Related Work}

Causal inference methods \cite{pearl1995causal, athey2016recursive, wager2018estimation,hill2011bayesian} are primarily shaped by two foundational frameworks: Structural Causal Models (SCM) by Pearl \cite{pearl2009causality}  and the Potential Outcomes (PO) framework by Rubin \cite{rubin1974estimating}. Recent literature emphasizes the importance of counterfactual analysis, which offers more direct interpretability in practical scenarios \cite{athey2017state, wachter2017counterfactual, verma2020counterfactual}. Wachter et al.~\cite{wachter2017counterfactual} introduce counterfactual explanations as a way to explain algorithmic decisions without revealing the model internals. Athey and Imbens \cite{athey2017state} synthesize advances in applied econometrics, highlighting counterfactual methods like difference-in-differences to assess policy impact. Although powerful, these methods are not user-centric towards the general public, which limits their ability to explore alternative scenarios.

Visual analytics has notably advanced in leveraging counterfactual reasoning to enhance interpretability and user interaction with causal inference models \cite{wang2015visual, wang2017visual, wang2022domino, kaul2021improving, guo2021vaine, guo2023causalvis, xie2020visual, wang2024beyond}. Wang et al.~\cite{wang2017visual} pioneered interactive visualization frameworks that support "what-if" simulations to elucidate causal relationships. Guo et al.~\cite{guo2023causalvis} provide interactive counterfactual analyses by enabling users to adjust confounders and refine cohorts. Wang et al.~\cite{wang2024beyond} integrate counterfactual reasoning into visual analytics guidance mechanisms to help users make more accurate causal inferences by exploring hypothetical changes in data subsets. This shift underscores the importance of user interaction in counterfactual-focused visual analytics.

To maximize the utility of visual analytics systems in highly heterogeneous contexts, one needs to adapt these analyses to local subgroups or contexts. LFPeers~\cite{sachdeva2023lfpeers} enables users to filter and investigate peer groups, uncovering divergent subgroup patterns. SUBPLEX~\cite{yuan2022subplex} advances cluster exploration and incorporates local model explanations across subpopulations. Similarly, CausalPrism~\cite{zhou2024causalprism} focuses on generating and interpreting counterfactual subgroups, enhancing user comprehension of treatment-effect heterogeneity.
XplainAct builds on insights from these prior methods, concentrating on supporting the analysis and interpretation of personalized interventions within subgroups.

\section{XplainAct}
During XplainAct’s design phase, we collaborated with NIH public health experts to understand their needs for visualizing and interacting with personalized interventions. Their feedback informed the following four design goals (DG):


\begin{itemize} [topsep=2pt, itemsep=2pt, parsep=2pt, leftmargin=15pt]
    \item \textbf{DG1: Specify intervention and estimate outcomes.} XplainAct should allow analysts to simulate interventions to a unit of interest and estimate potential outcomes under various treatments. 
    
    
    \item \textbf{DG2: Identify subgroups.} Since individual responses vary based on feature attributes (covariates), XplainAct should enable users to identify relevant subgroups to account for heterogeneity during intervention analysis.

    
    \item \textbf{DG3: Explain intervention results.} To support reasoning and build trust, the framework should elucidate how each attribute contributes to the outcome and reveal the sources of variation in predicted results.
    
    
    \item \textbf{DG4: Show feature-to-space relations.} XplainAct is motivated by contexts where units are distributed geographically, thereby requiring visual tools that can help users to recognize familiar spatial patterns and intuitively link descriptive features to locations.

\end{itemize}

XplainAct uses Python on the backend, employing Flask for handling HTTP requests, DoWhy for causal inference, and additional libraries for machine learning algorithms. The front end is built with JavaScript, using D3.js to create interactive visualizations.

\begin{figure}[t!]
\centering
\includegraphics[width=\linewidth]{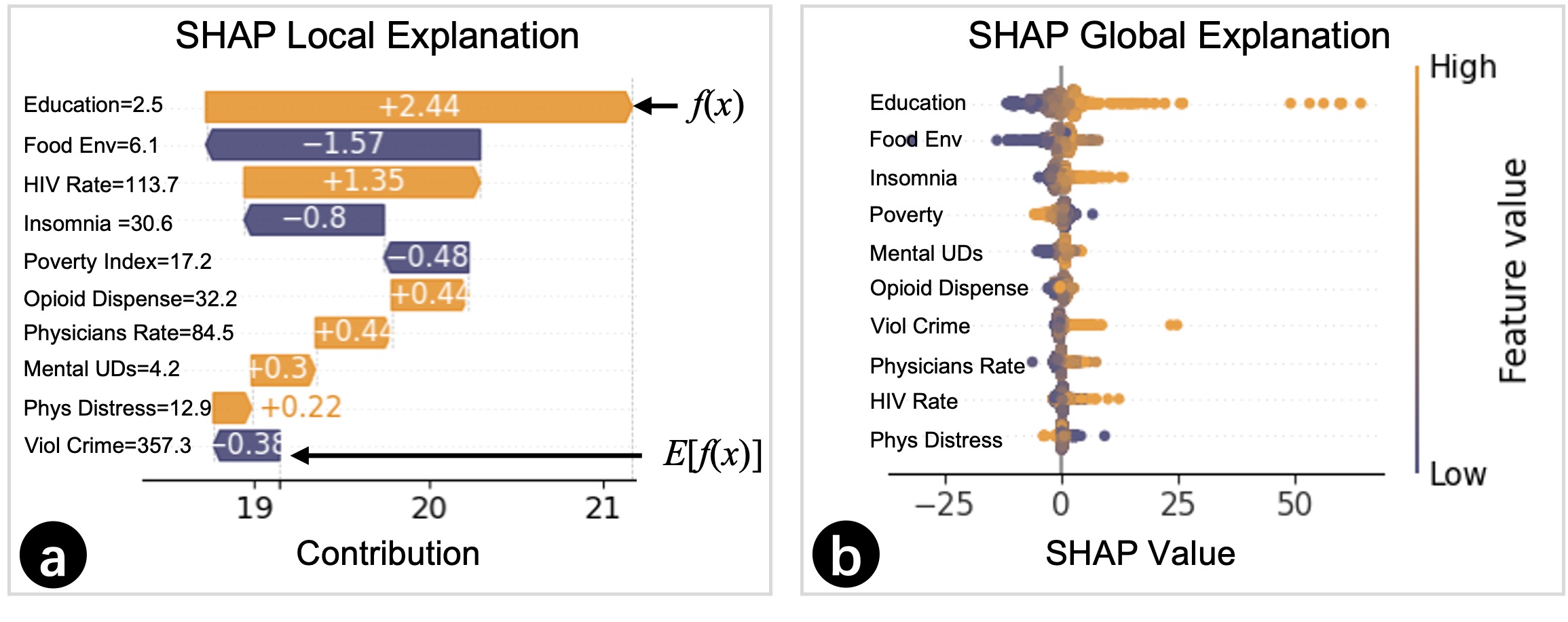}
\vspace{-15pt}
\caption{Explanations supported by SHAP. (a) Waterfall plot showing cumulative contributions from each feature. (b) Beeswarm plot showing SHAP values of each data point.}
\label{fig:explain}
\vspace{-10 pt}
\end{figure}

\subsection{Visual Components}
XplainAct has three views: the Choropleth Map view (A), the Explanation view (B), and the Subgroup view (C) (see \cref{fig:teaser}). 

The Choropleth Map view visualizes outcomes across geographic units using a diverging color scheme to differentiate polarities. In the case study of the opioid dataset, shades of purple represent higher opioid death rates, while green indicates lower ones. A neutral or medium outcome value is depicted in light yellow. Hovering over a geographic unit reveals a tooltip showing the unit's precise outcome value. Typically, users specify geographic units exhibiting extreme values, notable spatial patterns, or regions with which they have high familiarity \textbf{(DG4)}.


The Explanation view supports both local and global explainability for intervention results \textbf{(DG3)}. It integrates two explainable AI tools, LIME and SHAP, to provide complementary interpretations from distinctive perspectives. 

For local explainability using LIME (Local Interpretable Model-Agnostic Explanations \cite{ribeiro2016should}) (see \cref{fig:teaser}(B)), the interface displays a summary display showing the prediction and a diverging bar chart showing the contributions of every feature to the prediction. The display exhibits the predicted value of the outcome along with a variability interval derived from a set of perturbed samples \textbf{(DG1)}. In the bar chart, the bar length indicates the magnitude of a feature’s contribution, and its color signifies the sign of that contribution. Specifically, orange bars indicate features that increase the value of prediction (e.g., education index $>$ 1.92), while indigo bars indicate features that lower the value of prediction (e.g., poverty index $>$ 16.7). LIME approximates the model locally by learning a simple, interpretable model around the instance of interest—allowing users to understand which features are driving the prediction in that specific context.


Alternatively, users can switch to a waterfall plot generated using SHAP (SHapley Additive exPlanations \cite{lundberg2017unified}) for a different perspective on local explanations (see \cref{fig:explain}(a)). A SHAP waterfall plot explains a single prediction by decomposing it into additive contributions from each input feature. Starting from the model’s baseline (i.e., the average prediction across the dataset), the plot shows how each feature pushes the prediction higher or lower, using bars whose lengths represent the magnitude of the contribution. These contributions are computed based on Shapley values, which quantify the average marginal effect of each feature across all possible subsets of features. The resulting plot highlights how much each feature contributed to the final prediction and in what direction, using a consistent color scheme to distinguish positive and negative influences.


Global explainability is visualized using SHAP beeswarm plots (see \cref{fig:explain}(b)), where each data (i.e., county) is shown as a scatter point. The horizontal axis represents the SHAP value, indicating how much a feature contributes to the model output for each instance. Features are listed vertically, ordered by their overall importance (i.e., the average absolute SHAP value across all instances). Each point is colored by the corresponding feature value, transitioning from low (indigo) to high (orange), thus revealing how different value ranges influence predictions. The beeswarm plot combines feature importance with value distribution in a single view, for both granular and high-level insight into model behavior. 

The Subgroup view has two modes: \textit{profile mode} and \textit{intervention mode}. The profile mode features a slider group and a parallel coordinates plot (PCP) (see \cref{fig:teaser}(c, d)). Slider values, reflecting the feature values of the unit $x$, correspond to a prominent red polyline in the PC plot. Additionally, the ten nearest  neighbors of $x$, based on feature attributes, appear as thinner blue polylines. In intervention mode (see \cref{fig:intervene}), users can adjust the subgroup defined as $x$'s $N$ nearest neighbors via the slider on the far left \textbf{(DG2)}. The PCP is enhanced with translucent red rectangles on each axis to indicate the value ranges defining the adjusted subgroup. In this mode, only the polyline of the selected unit (shown in red) and its counterfactual polyline (shown in blue) are displayed, with the intervention attribute highlighted by a blue axis title.



\begin{figure}[t!]
\centering
\includegraphics[width=0.8\linewidth]{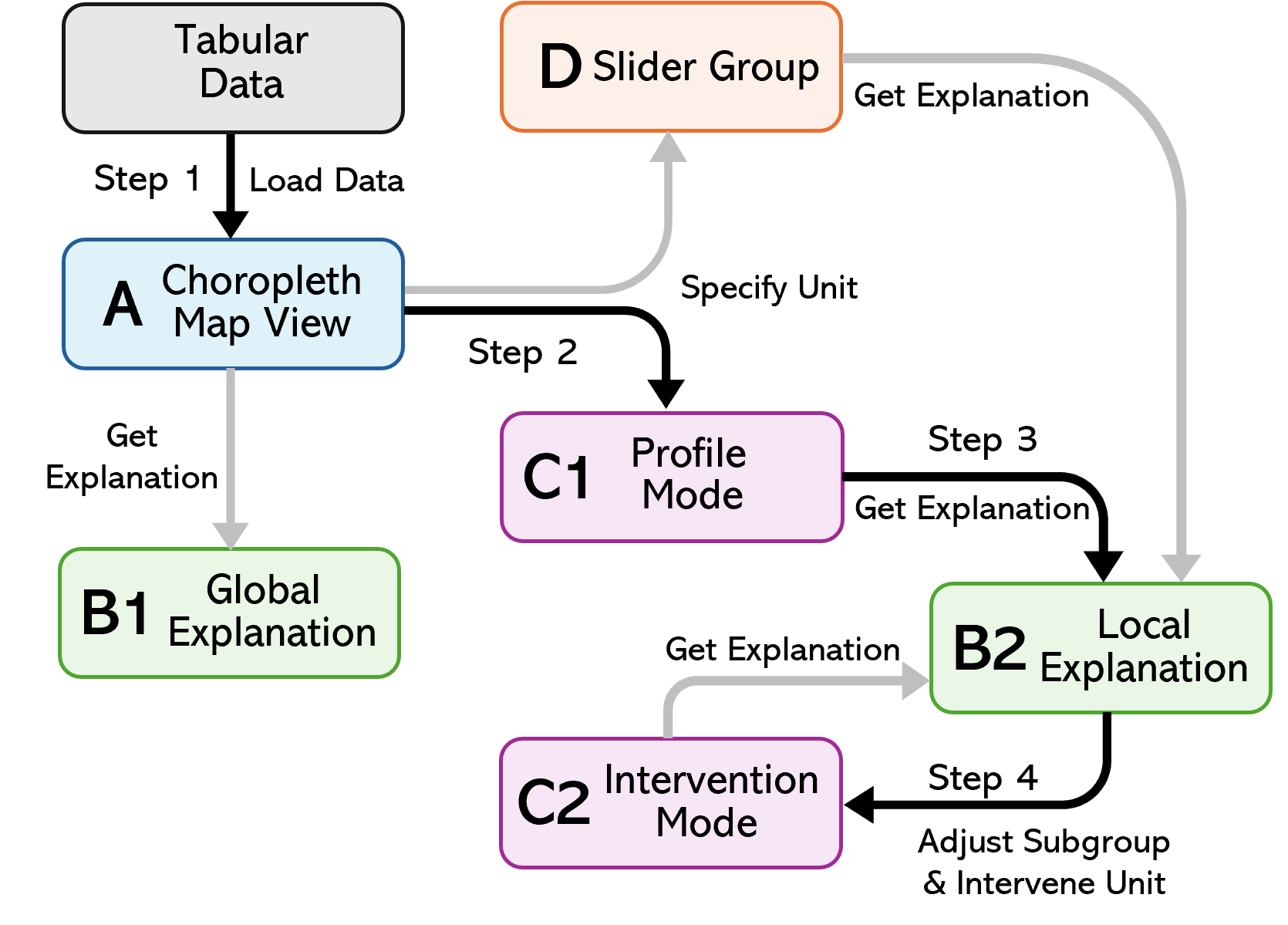}
\vspace{-10pt}
\caption{Workflow of XplainAct for analyzing personalized interventions. Each block represents a view, and each edge denotes an interaction between views.}
\label{fig:workflow}
\vspace{-10 pt}
\end{figure}

\subsection{Interactions}
\cref{fig:workflow} depicts the iterative process of XplainAct, operationalized through the dashboard interface shown in \cref{fig:teaser}. Starting with a tabular dataset, the choropleth map visualizes the spatial distribution of the outcome variable, helping analysts identify areas of interest (Step 1). When a unit of interest (county $x$) is selected from the map, the map highlights $x$ with a thicker red outline and shows its peers based on feature similarity. This is reflected in the profile mode of the subgroup view, where the sliders and the PCP are updated accordingly (Step~2). Next, analysts can adjust attribute values with sliders or request local model explanations for \( x \) by clicking the \textit{Get Explanation} button. The resulting explanations reveal how individual attributes affect the predicted value of the outcome, guiding subsequent intervention strategies (Step~3).


In Step 4, informed by these insights, analysts switch to intervention mode and simulate a desired intervention by clicking an attribute value on the targeted axis through the PCP (see \cref{fig:intervene}). The system simulates the impact of the selected intervention -- the treatment -- and demonstrates the changes in related attributes through the generated counterfactual unit $x'$ represented by the blue polyline \textbf{(DG1)}. In this process, the values of attributes independent of the treatment remain unchanged -- only the values for its dependent attributes are affected, reflecting realistic intervention outcomes \textbf{(DG3)}. Meanwhile, analysts can also adjust the subgroup by modifying the number of nearest neighbors ($N$) for unit $x$ \textbf{(DG2)}.


Internally, XplainAct constructs a subgroup by identifying similar units using efficient similarity searches (Locality Sensitive Hashing~\cite{gionis1999similarity}), then parameterizes a localized causal model based on the selected peers. This model is capable of predicting counterfactual outcomes~\cite{pearl20217} within the subgroup. The underlying structure of this local causal model is static and is either constructed by a domain expert beforehand or derived from causal discovery tools \cite{spirtes2000causation, chickering2002optimal, zhang2024causalchat}. Further details about the causal modeling approach are provided in the supplementary material.

By engaging iteratively with Steps 2–4, analysts can systematically explore personalized interventions, estimate outcomes, and generate explanations. Once a unit is specified (Step 2), the feature importance shown in the explanation view provides insights that help analysts identify which feature attributes to target for intervention (Step 3). Analysts can then refine the subgroup and apply interventions accordingly (Step 4).

\section{Case Studies}
We demonstrate the capabilities of XplainAct by presenting two usage scenarios that employ real-world datasets. Visuals for these two investigations can be found in the supplementary material.

\textbf{The Opioid Death Dataset} focuses on the relation between opioid-related deaths and social factors. It combines 10 key socioeconomic attributes sourced from the County Health Ranking database \cite{health2023ranking} with opioid death data from the CDC WONDER database \cite{cdc2021opioid}. It covers over 3,000 US counties. In the dataset, each county is characterized by 11 attributes: food environment index, primary care physicians rate, violent crime rate, HIV prevalence rate, education index, poverty index, percent insufficient sleep, average mental unhealthy days, percent frequent physical distress, opioid dispensing rate, and opioid death rate. Opioid death rate is the outcome variable, and other variables are treatment variables.

\textbf{The Presidential Election Dataset} centers on the correlation between voting inclination and socioeconomic drivers. The dataset includes 9 carefully selected socioeconomic attributes from a 2016 US election dataset \cite{hamner_2016_us_election} and each data point is a US county. The attributes are: percent rural population, percent minority population, percent population physically inactive, percent household own home, number of unemployment, percent population age 65 and older, violent crimes per 100k, education index, and percent population Black. The outcome variable is vote percentage difference with positive value leaning towards Party B's candidate and negative value leaning towards Party A's candidate. 

\subsection{Case Study 1: Opioid-Related Death Dataset}
We follow Taylor, a health-conscious citizen who is deeply concerned about the opioid crisis in her home of Boone County, West Virginia. She turns to XplainAct in hopes of understanding why her county is struggling with high opioid mortality and identifying actionable steps that could help reduce the death rate. 

She begins by analyzing the choropleth map to understand how opioid-related deaths vary across counties. By exploring several high-death-rate areas (shaded in purple) and low-death-rate areas (shaded in green), she examines their attribute distributions in the PCP. This exploration gives her an overview of how different attributes appear to correlate with opioid mortality, helping her form initial hypotheses about potential risk factors.

Next, she decides to focus on Boone County and its cluster by selecting it on the choropleth map (see \cref{fig:teaser} for a similar analysis). The map highlights Boone and its peer counties, and the PCP immediately reveals several concerning patterns shared across the cluster: high levels of insufficient sleep, a large number of mentally unhealthy days, and low education levels. These trends stand out as potential contributing factors to the cluster’s alarmingly high opioid death rate.

To explore potential mitigating strategies, Taylor switches to the intervention mode of the PCP. In this mode, she interacts with the model to simulate the effects of various hypothetical interventions. After experimenting with different attributes, she discovers that the \textit{opioid death rate} can be significantly reduced by lowering the \textit{number of mentally unhealthy days}. Further exploration reveals that decreasing the \textit{percent of insufficient sleep} also reduces the \textit{number of mentally unhealthy days}, leading to an additional drop in the \textit{opioid death rate}. This causal chain offers actionable insights for designing targeted interventions to address the crisis in her community.

To better understand Boone County’s situation, Taylor turns to the explanation view. The waterfall plot generated using SHAP highlights key factors contributing to the high \textit{opioid death rate}, including high \textit{percent of insufficient sleep}, a high \textit{number of mentally unhealthy days}, elevated \textit{HIV prevalence rate}, and a high \textit{violent crime rate}. The first two attributes confirm the trends she observed earlier, while all four collectively reveal the underlying challenges contributing to the opioid crisis in Boone County and its peers.

Based on these findings, Taylor recommends that Boone County promote better sleep habits through public health and workplace initiatives, expand access to health services, and address broader issues like crime and education. She also suggests that other counties with similar patterns could apply these priorities to reduce opioid-related challenges.



\begin{figure}[t!]
\centering
\includegraphics[width=\linewidth]{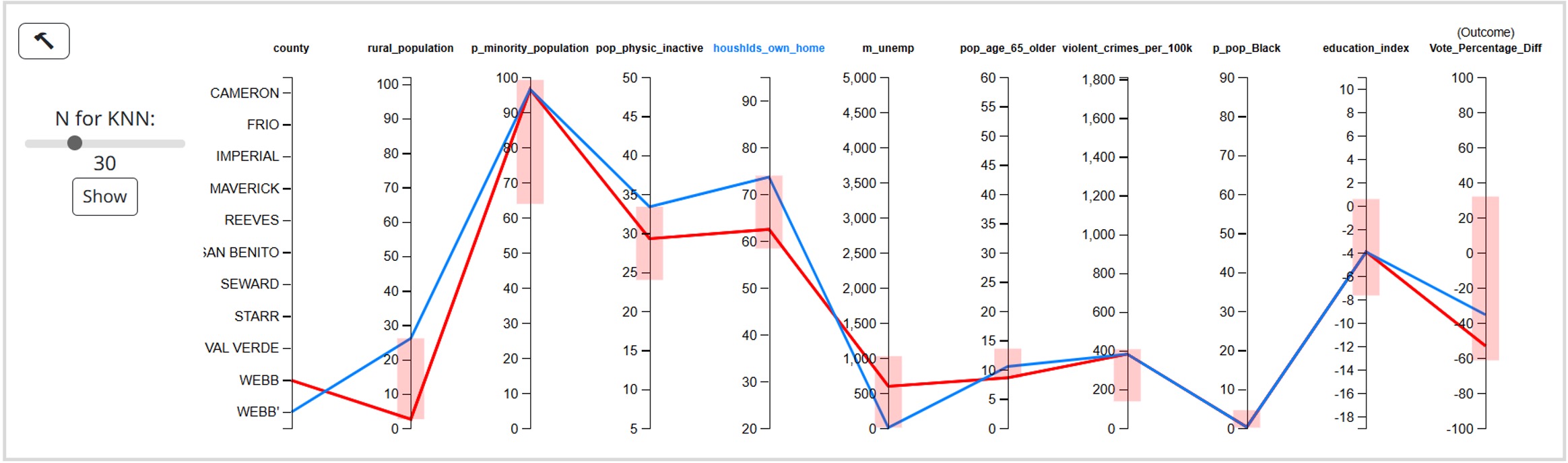}
\vspace{-15pt}
\caption{Bob simulates an intervention on increasing household homeownership and observes an advantage for party B.}
\label{fig:intervene}
\vspace{-10 pt}
\end{figure}

\subsection{Case Study 2: Presidential Election Dataset}

In this study, we follow Bob, a campaign staffer of Party B, as he uses XplainAct to investigate why Party A secured a 53\% lead in Webb County, Texas. His goal is to uncover factors influencing voter turnout and explore strategies to boost Party B's support by mobilizing the “right” voters in future elections.

Bob begins his exploration by examining the choropleth map, which provides a geographic overview of election results across the country. In Texas, most counties lean Party B, except for South Texas, where Webb County is located. Upon clicking Webb County, he observes that counties with similar attributes are mostly concentrated in South Texas and California, nearly all are Party A leaning. He then proceeds to the intervention mode, where he adjusts the number of neighbors considered for Webb County. This helps him assess whether Webb’s voting trend reflects a broader regional pattern or represents an outlier.

Meanwhile, the pattern shown in the PCP draws Bob’s attention. The subgroup of Webb County is characterized by a low \textit{rural population}, a low \textit{percentage of residents aged 65 and older}, and a high \textit{percentage of minority residents}. It also shows relatively low values for \textit{unemployment rate}, \textit{violent crime rate}, \textit{percentage of Black population}, and \textit{education index}. Bob agrees that many of these attributes align with his understanding of Webb County. However, some characteristics do not match common stereotypes of a Party A leaning county, prompting his curiosity about how each attribute influences the vote difference.

Bob then turns to the explanation view to examine the LIME and SHAP plots for Webb County. These plots decompose the impact of each attribute on the county’s vote difference. Bob discovers that both LIME and SHAP identify the percentage of minority population as the most significant factor contributing to Webb's Party A lean. Specifically, the SHAP waterfall plot shows that Webb County’s 95\% minority population alone contributes a -72.48
shift from 
the national average vote difference of 32.01.  
This places Webb as a Party A stronghold, consistent with common expectations. Other attributes—such as textit{low rural population} and moderately low \textit{homeownership rate}—further reinforce this alignment.

Conversely, Webb also exhibits characteristics typically associated with Party B's support, such as a low \textit{percentage of Black residents} and a moderately low \textit{education index}. To his surprise, Bob notices that within Webb’s subgroup, a low percentage of elderly residents pushes the county further toward the Party B side. This goes against the common perception that Party B performs strongly among older voters and suggests that Webb County may have a unique age-related voting pattern.

With this knowledge in mind, Bob now seeks to explore potential strategies for improving Party B's performance. He toggles the PCP to intervention mode, where the interactive interface allows him to simulate changes and observe their impact on the vote difference. First, he adjusts the percentage of minority population and finds that decreasing this attribute predicts an increase in Party B's votes. He also explores the effect of increasing the proportion of homeowners (see \cref{fig:intervene}), which results in a positive shift toward Party B. While population structure cannot be directly changed, the intervention suggests that policies or campaigns targeting homeowners and encouraging them to vote could be an effective way to gain traction for Party B in Webb County. Armed with these insights, Bob is prepared to recommend targeted strategies to improve Party B's performance in future elections.


\section{Limitations and Future Work}
XplainAct performs well on structured datasets with a predefined outcome and covariates, typically containing 15 or fewer attributes at the county or similar geographic settings. Yet, current pipeline struggles with large datasets containing hundreds of attributes. This limitation also affects the representation and interaction with visual tools like the PCP. To mitigate this, future work can incorporate attribute grouping or dimensionality reduction to manage complexity.

Another limitation is the flexibility of subgroup definition. In the current pipeline, subgroups are merely decided by the unit's nearest neighbors. This highlights the need for more flexible subgroup exploration and identification methods.

Moreover, the use of uniform causal structures across subgroups, though it simplifies computation, obscures meaningful subgroup-specific causal differences. XplainAct omits causal graph visuals due to the trade-off between transparency and usability. While graphical causal models can clarify the decision-making process—such as showing how independent variables influence a collider—but they also raise the entry barrier, as users may need prior knowledge of data causality. Future work may integrate subgroup-specific causal graphs to support users with relevant expertise.



\section{Conclusion}
In this paper, we presented XplainAct, a visual analytics framework for exploring individual-level interventions within subgroups.

Despite current limitations, XplainAct advances visual frameworks for interpreting decision-making in complex, heterogeneous scenarios. This advancement ensures both practical relevance and enhanced analytical power in real-world applications.


\acknowledgments{
This research was funded in part by CDC Cooperative Agreement 1 NH25PS005202-01-00 and American Public Health Association (APHA) award  \# 2023-0004.}

\bibliographystyle{abbrv-doi}

\bibliography{template}
\end{document}